\pdfoutput=1

\documentclass[english,aps,prstper,reprint,showpacs,titlepage,longbibliography,superscriptaddress]{revtex4-1}   

\usepackage[T1]{fontenc}	% should generally be included for better accented-word behavior
\usepackage[latin9]{inputenc}	% should generally be included for better accent behavior
\usepackage{geometry}		% for controlling page margins
\usepackage{graphicx}
\usepackage{times}

\usepackage{hyperref}  
\hypersetup{colorlinks=true,urlcolor=blue,citecolor=blue,linkcolor=blue}   
\usepackage{enumitem}          % package for handling list formatting
\setlist{nosep}                 % Tightest spacing for lists. `noitemsep` is more relaxed

\usepackage[caption=false]{subfig}

\begin{document}

\begin{titlepage}

  \title{Student reasoning about sources of experimental measurement uncertainty in quantum versus classical mechanics}

  \author{Emily M. Stump}
  \affiliation{Laboratory of Atomic and Solid State Physics, Cornell University, 245 East Avenue, Ithaca, NY,
    14853} 
  \author{Courtney L. White}
  \author{Gina Passante}
  \affiliation{Department of Physics, California State University Fullerton, 800 N. State College Blvd., Fullerton, CA,
    92831} 
   \author{N. G. Holmes}
   \affiliation{Laboratory of Atomic and Solid State Physics, Cornell University, 245 East Avenue, Ithaca, NY,
    14853} 

  \begin{abstract}
    Measurement uncertainty and experimental error are important concepts taught in undergraduate physics laboratories. Although student ideas about error and uncertainty in introductory classical mechanics lab experiments have been studied extensively, there is relatively limited research on student thinking about experimental measurement uncertainty in quantum mechanics. In this work, we used semi-structured interviews to study advanced physics students' interpretations of fictitious data distributions from two common undergraduate laboratory experiments in quantum mechanics and one in classical mechanics. To analyze these interpretations, we developed a coding scheme that classifies student responses based on what factors they believe create uncertainty and differentiates between different types of uncertainty (e.g. imprecision, inaccuracy). We found that participants in our study expressed a variety of ideas about measurement uncertainty that varied with the context (classical/quantum) and the type of uncertainty.\clearpage
  \end{abstract}

  \maketitle
\end{titlepage}

\section{Introduction}

One of the goals of an undergraduate physics education is that students understand the process of experimental scientific inquiry. To evaluate experimental results and claims, one must be able to assess the quality of the methods and data~\cite{WiemanCTA}. However, research has consistently shown that many introductory physics students hold ideas about measurement and uncertainty that differ from experts'~\cite{buffler2001development}. 

One source of confusion lies in terminology. Physicists typically use the term ``error'' to be synonymous with ``uncertainty,'' which may refer to what is unknown about a measurement or the degree to which a measurement is limited. The colloquial use of error, however, is to describe a mistake. Research has found that many students interpret the physicist term ``error'' in this colloquial sense in classical mechanics experiments~\cite{Evangelinos2002, HolmesBFY2015}. This line of thinking has been observed in students' broader understanding of measurement. For example, students may expect that all errors (and thus uncertainty) can be eliminated by experts~\cite{Evangelinos2002}. 

Terminology poses a similar issue for measurement and uncertainty in quantum mechanics. In quantum mechanics, uncertainty may refer to ``the degree to which measurements on a microscopic scale are intrinsically non-reproducible''~\cite[p. 439]{Johnston1998}. This distinct definition of uncertainty, conflated with students' confusion about uncertainty in classical mechanics, can lead to further complications in student understanding. 

To our knowledge, there is relatively limited research on student thinking about experimental measurement uncertainty in quantum mechanics. Several studies have shown that students draw on classical-physics reasoning to explain quantum-mechanical systems~\cite{passante2015examining, Kalkanis2003, Singh2001}. For example, one study found that many third-year undergraduate students interpreted quantum-mechanical uncertainty as a measure of what is unknown or limited~\cite{Johnston1998}. However, a recent study found that students' epistemological views of classical and quantum mechanics are distinct~\cite{Dreyfus2019}. 

In this study, we aimed to capture student ideas about sources of measurement uncertainty in both classical and quantum contexts. To classify these ideas, we drew on the Modeling Framework for Experimental Physics, which models experimental physics as generating, comparing, and refining models of physical and measurement systems~\cite{zwickl2015model}. Because a model is a simplified representation of the physical world~\cite{Brewe2008}, principles of uncertainty will be intrinsic to the modeling process. We can thus use the modeling framework to characterize student thinking in classical and quantum contexts within the same coding scheme. Research indicates that students draw on different forms of reasoning to answer different questions~\cite{Leach2000}, so it is plausible that they would use different reasoning in these different contexts. However, other work found that students do not consider theoretical models when interpreting data~\cite{Ryder2000}, so when shown actual experimental data, students may ignore the theoretical context.

Previous work had developed an emergent coding scheme to analyze a subset of questions and scenarios from the interview data used in this study and found that students identified very different sources of uncertainty in the two contexts~\cite{Stein2019}. Here, we expand that work to evaluate whether this polarization might have resulted from the emergent coding scheme rather than underlying differences in student thinking. We also expand that work by probing student thinking in all parts of the interviews and disentangling references to precision and accuracy within the contexts. In particular, we sought to determine whether students attribute sources of uncertainty (generically, in reference to precision, or in reference to accuracy) to limitations or principles of the physical or measurement models, and how those attributions vary by context. 

\section{Methods}

\begin{table*}[htbp] % placed here to be closer in text to coding scheme description
  \caption{The coding scheme used to characterize student ideas about sources of measurement uncertainty, shown in relationship to the Modeling Framework for Experimental Physics~\cite{zwickl2015model}. The term distribution here refers to any spread or variability in the data.\label{codes}}
  \begin{ruledtabular}
    \begin{tabular}{p{.1\textwidth}p{.42\textwidth}p{.42\textwidth}}
      & \textbf{Measurement model} & \textbf{Physical model}\\
      \hline
      \textbf{Limitations} & Distribution due to mistakes caused by either the experimenters or measuring device. (ML) &  Distribution due to invalid assumptions or variables not accounted for in the model. (PL)\\
      \textbf{Principles} & Distribution due to inherent limitations in the measurement apparatus, i.e. measurement devices cannot be infinitely precise. (MP) & Distribution due to inherent features of the physical model, such as the uncertainty principle, wave functions, or unknowable ``hidden variables.'' (PP)
    \end{tabular}
  \end{ruledtabular}
\end{table*}

\subsection{Participants and data collection}

We used semi-structured interviews to probe student thinking about measurement uncertainty. Nineteen students were interviewed for this study, including 17 undergraduate and two graduate students. Participants were recruited from two institutions: one a research-intensive, private, and PhD-granting institution and the other a public, master's-granting, and Hispanic-serving institution. All participants had either previously taken a quantum mechanics course or were currently taking one at the time of the interview. Interviews took place in the second half of the spring semester.

The interviews focused on measurement uncertainty in the context of three experiments: one classical and two quantum-mechanical. For each experiment, the participant was presented with a histogram of fictitious data hypothetically collected by students conducting the experiment. 

\begin{figure}
  \subfloat[]{\includegraphics[width=.238\textwidth]{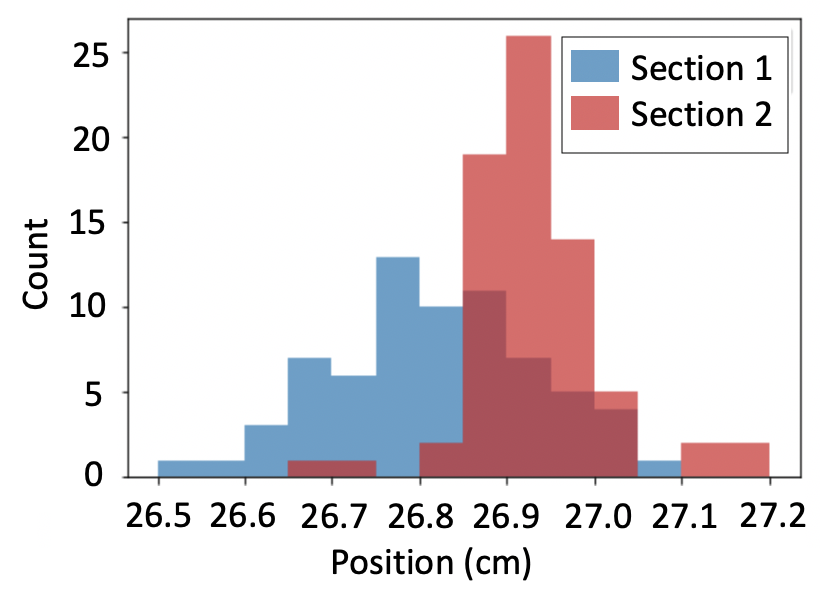}\label{ball_drop_hist}}
  \subfloat[]{\includegraphics[width=.238\textwidth]{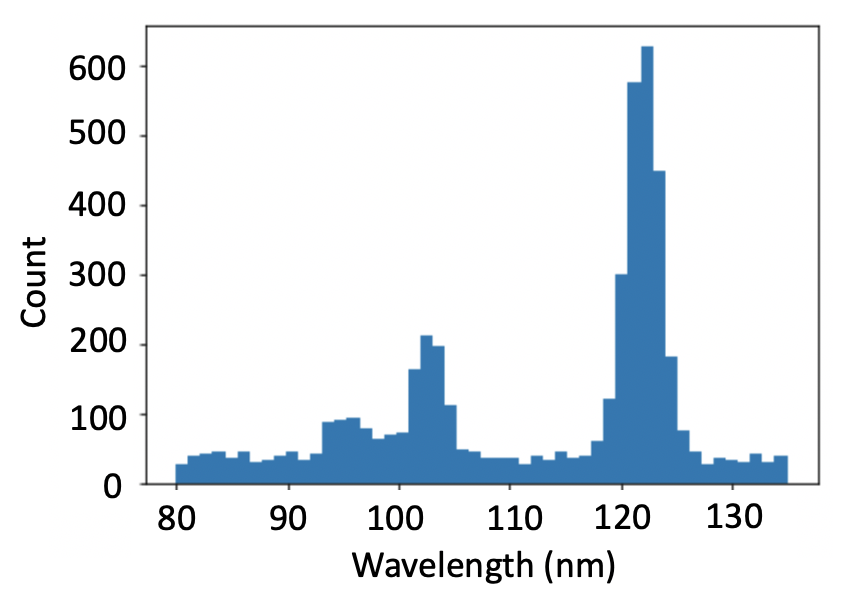}\label{spectrum_hist}}
  \caption{Histograms of (a) ball-drop and (b) spectrum data presented to participants during the interview. For the single-slit experiment, participants were shown the histograms in (a) with the x-axis altered.}\label{fig:data}
\end{figure}

The classical mechanics scenario was a ball-drop experiment adapted from the Physics Measurement Questionnaire~\cite{buffler2001development}. In the scenario, 63 students released a ball from a ramp on a table, measured the landing position, and produced the blue histogram in Fig.~\ref{ball_drop_hist}. Participants were asked to explain the shape and source of the distribution. They were then asked to compare the data to the result an instructor would obtain if they measured the release height and calculated the predicted landing position. Finally, participants were asked to compare the original data to the red histogram in Fig.~\ref{ball_drop_hist}, a data set hypothetically collected by another group of students. 

The analogous quantum-mechanical scenario was a single-photon single-slit experiment that used the same data sets as the classical scenario (Fig.~\ref{ball_drop_hist}). In this scenario, the histograms represented the position of 63 single-photon measurements on a screen. The interview questions had the same structure as for the ball-drop scenario, with students discussing the shape of the histogram, a comparison to a calculation of the diffraction pattern from a measurement of the slit width, and a comparison between two sets of student data.

Between the experiments above, students were asked about another quantum-mechanical experiment: measurements of the wavelength of light emitted from an object (fictional data shown in Fig.~\ref{spectrum_hist}). Participants were asked similar questions about the shape of the histogram as above but were not asked to compare to another data set or a theoretical calculation.

In addition to the specific scenarios, participants were asked about measurement uncertainty in classical and quantum mechanics without reference to particular experiments: ``\textit{What comes to mind when you think about measurement uncertainty in classical (quantum) mechanics?}'' These questions were posed twice: once between the spectrum and single-slit experiments and once at the end of the interview.

\subsection{Coding scheme}

The codes for student-identified sources of uncertainty, chosen to align with the Modeling Framework for Experimental Physics~\cite{zwickl2015model} and shown in Table~\ref{codes}, were Measurement model Limitations (ML), Measurement model Principles (MP), Physical model Limitations (PL), and Physical model Principles (PP). Measurement model limitations included human error (as in Ref.~\cite{Stein2019}) or inconsistencies between trials, as exemplified by ``\textit{if the measuring stick wasn't exactly like along the same line or axis every time where, or if it was slightly shifted.}'' Measurement model principles referred to measuring devices not being infinitely precise, for example, ``\textit{So it would just be a bunch of results at the same position like plus minus whatever the, what's it called, whatever precision is a yard stick goes.}'' Physical model limitations included confounding variables, such as friction, air resistance, and background light, and variables missing from the physical model. Example participant statements included ``\textit{There might be like 
the path that the ball takes would be different and there might be like different friction coefficients along the ramp}'' and ``\textit{Well that probably means that there are other variables going on. You know, that the model is not entirely accounting for everything.}'' Physical model principles included sources of variability inherent to the system, particularly due to the probabilistic nature of quantum mechanics: ``\textit{Uh, always talk about the Heisenberg uncertainty principle [...] And, uh, we're taught to think of it as a, um, intrinsic property of the theory of how nature, how it works.}'' In the spectrum experiment, physical model principles also included reasons for the multiple peaks in the histogram, such as, ``\textit{So like the reason why there could be multiple ones are because [...] there's I guess an increments of energy levels that you can go.}'' 

We coded any statement about the sources of measurement variability throughout the interview. Some statements were too vague to categorize, but otherwise all statements about sources of uncertainty were codeable based on this scheme. Rather than count unique instances of each code for each student, we identified \emph{whether} a student described each type of uncertainty source in Table~\ref{codes}, and then analyzed the fraction of students that drew on each type throughout the interview.

Participants' responses were also coded for the \textit{type} of uncertainty described. Statements addressing the source of spread, imprecision, or variability were coded as \emph{precision}. Statements about sources of systematic error or inaccuracy were coded as \emph{accuracy}. Because statements about accuracy appeared almost exclusively in comparisons of two data sets or comparisons of a data set to the value calculated from theory, the accuracy code was further divided into two codes designating the context: \emph{data versus data} and \emph{data versus theory}. Statements in response to the question ``\textit{What comes to mind when you think about measurement uncertainty in classical (quantum) mechanics?}'' were coded as \emph{generic measurement uncertainty}. Participants responded similarly both times this question was posed, so we collapsed the instances together.

The first and second authors independently coded the transcripts of all nineteen interviews. We calculated inter-rater reliability through Cohen's kappa for each  source code across all types of uncertainty. The Cohen's kappa values were 0.857 for measurement limitations, 0.751 for measurement principles, 0.748 for physical limitations, 0.848 for physical principles. All coding disagreements were discussed and resolved.

\section{Results}

We present our results by each type of uncertainty and compare participants' responses between the classical and quantum contexts. We found, overall, that participants described different sources of uncertainty between the contexts. 

\subsection{Generic measurement uncertainty}
\label{uncertainty}

\begin{figure}
  \centering
  \subfloat[]{\includegraphics[width=.45\textwidth]{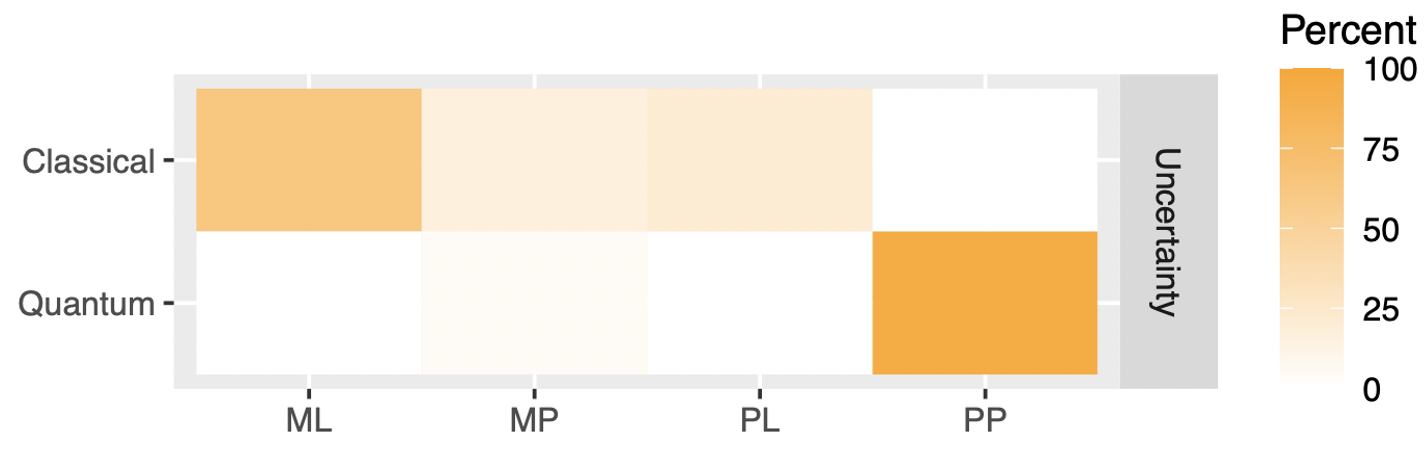}\label{heatmap_uncertainty}}\\\centering
  \subfloat[]{\includegraphics[width=.45\textwidth]{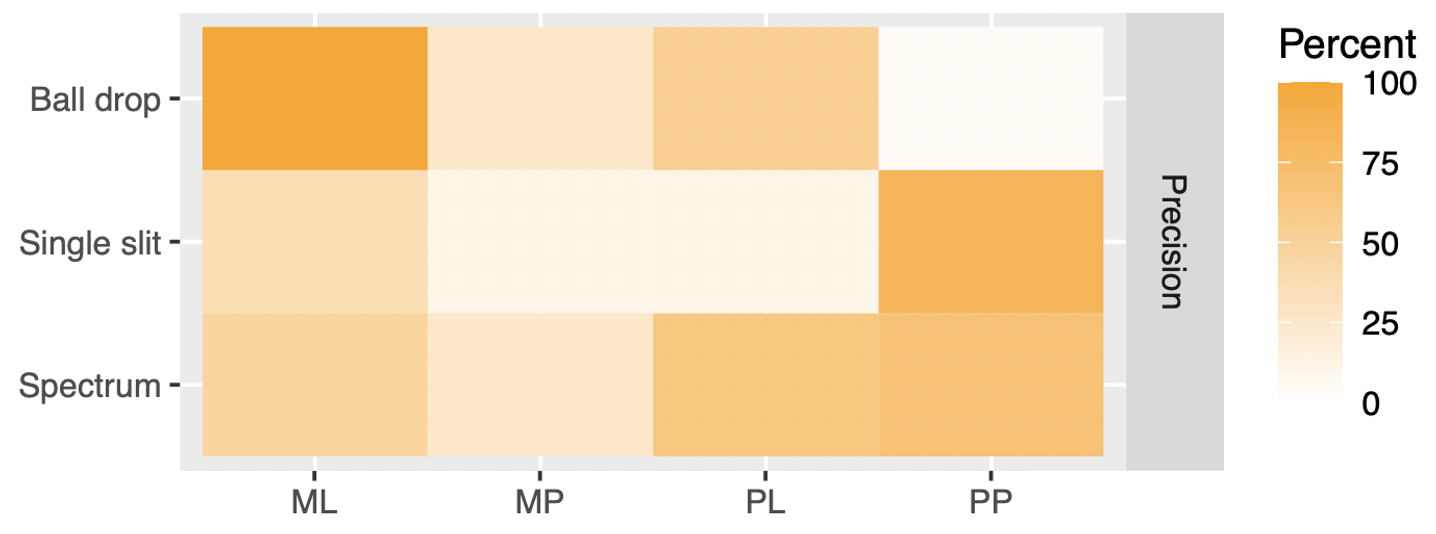}\label{heatmap_precision}}\\
  \centering
  \subfloat[]{\includegraphics[width=.45\textwidth]{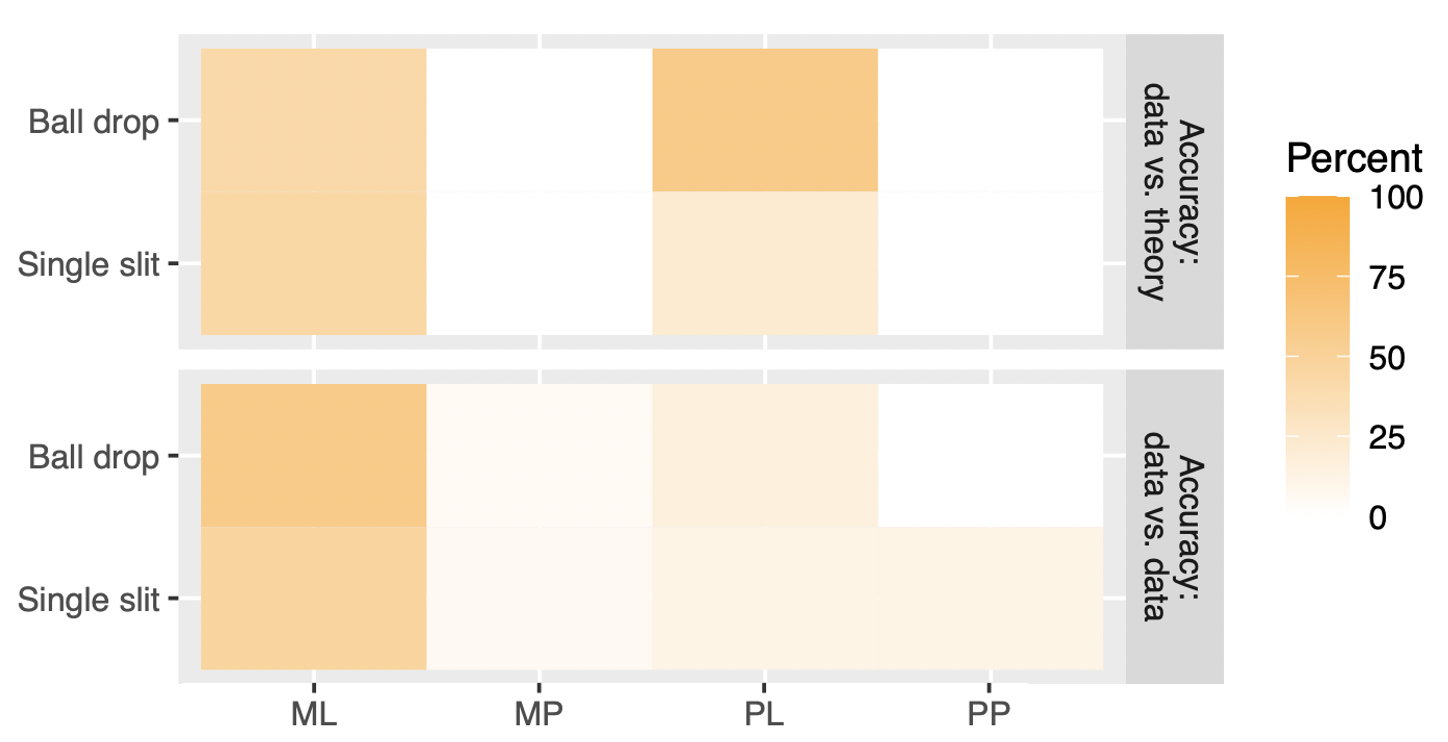}\label{heatmap_accuracy}}\\
  \caption{Sources of (a) generic measurement uncertainty, (b) precision, and (c) accuracy mentioned by interview participants. Note that the source codes are not mutually exclusive; thus, the rows do not add up to 100\%.}\label{fig:data}
\end{figure}

Figure~\ref{heatmap_uncertainty} shows the percent of participants who identified each type of source in response to the questions about generic uncertainty in classical and quantum mechanics. In \textit{classical mechanics}, most participants focused on measurement limitations, with fewer than half identifying measurement principles or physical limitations and none mentioning physical principles. In contrast, when asked about measurement uncertainty in \textit{quantum mechanics}, all but one of the participants attributed uncertainty to physical principles. Participants claimed that measurement uncertainty is inherent in the theory of quantum mechanics, with many explicitly referencing the Heisenberg uncertainty principle.

\subsection{Precision}
\label{precision}

As with generic measurement uncertainty, the types of responses about precision, shown in Fig.~\ref{heatmap_precision}, varied between the contexts, but with new subtleties.

When asked about the ball-drop experiment (\textit{classical}), all participants mentioned some form of measurement limitations. Compared with the generic uncertainty question, more participants also referenced physical limitations and measurement principles when considering the fictional data set. This suggest that participants considered more varied sources of uncertainty in response to the data than the generic question. 

For the single-slit experiment (\textit{quantum}), students' responses were also more varied when considering the data set than when considering measurement uncertainty in quantum mechanics generically. While sources of generic uncertainty in quantum mechanics were almost exclusively restricted to physical principles, explanations of the single-slit experiment data included physical principles (more than half the participants) and measurement limitations (fewer than half the participants), as well as a couple mentions of measurement principles and physical limitations. 

For the spectrum experiment (\textit{quantum}), participants' responses did not match their reasoning about either the ball-drop or single-slit experiments, and explanations were coded across all four categories. More than half the participants mentioned physical principles, alluding to the multiple peaks in the histogram as indicative of intrinsic properties of the material studied. However, more than half also cited physical limitations, with an emphasis on background light as a confounding variable. Several participants also referenced measurement limitations and measurement principles.

\subsection{Accuracy}

Participants' responses regarding accuracy are shown in Fig. \ref{heatmap_accuracy}, separated by whether students were referring to the comparison between the two student data sets or between the student data set and the value calculated from theory.

Responses comparing data to theory were similar between the ball-drop (\textit{classical}) and single-slit (\textit{quantum}) contexts, with participants drawing only on measurement limitations and physical limitations. In both cases, measurement limitations were typically attributed to some form of human error. In the ball-drop experiment (\textit{classical}, over half the participants mentioned physical limitations through confounding variables missing from the theory, such as friction or air resistance. In the single-slit experiment (\textit{quantum}), only four participants mentioned physical limitations, either suggesting that the slit was too wide to produce a diffraction pattern, or that the hypothetical theory calculation would not account for all variables present in the experiment. These similarities starkly contrast the distinct responses in the classical and quantum contexts in the previous sections.

In comparing two data sets, participants most frequently mentioned measurement limitations for both experiments, with physical limitations the next most common source. This suggests that participants attributed discrepancies between the data sets primarily to limitations, and that measurement limitations were more likely than physical limitations.

\section{Discussion}

In this study, we analyzed interviews with middle- and upper-division physics majors about sources of measurement uncertainty in classical and quantum-mechanical scenarios. The participants discussed measurement uncertainty differently in the two contexts, most evidently in their descriptions of generic measurement uncertainty. Consistent with the  previous analysis of this data~\cite{Stein2019}, we found that participants attributed generic classical uncertainty to a variety of measurement limitations, measurement principles, and physical limitations but almost exclusively discussed generic quantum uncertainty as inherent physical principles. 

These views of measurement uncertainty were largely borne out in participants' discussion of precision in the ball-drop and single-slit data. However, more participants drew on additional types of sources of measurement uncertainty when discussing the single-slit experiment than when describing quantum uncertainty generically, and with higher frequency than the previous analysis found~\cite{Stein2019}. This is consistent with other work indicating that students draw on different forms of reasoning when responding to a specific scenario as compared to a decontextualized question~\cite{Leach2000}.

Notably, the distinctions between classical and quantum contexts did not bear out in discussion of the spectrum experiment, where students drew almost equally on all type of uncertainty sources. Interestingly, a common explanation given for the spread in the spectrum data was background light, but no students drew on this idea in the single-slit context. One possible conclusion is that the participants did not think about the spectrum experiment as purely quantum-mechanical. This would align with previous work showing that many university students at a variety of educational levels apply mixed quantum and classical concepts in thinking about models that explain atomic spectra~\cite{Kalkanis2003,Mueller1999,Testa2019}. Indeed, participants drew on measurement principles and physical limitations in the spectrum experiment with similar frequency to the ball-drop experiment and on measurement limitations and physical principles similarly to the single-slit experiment. 

Alternatively, the pattern in responses could be the result of cueing from the interview questions~\cite{Russ2012}. The spectrum experiment questions were posed immediately after the ball-drop experiment and before quantum mechanics was mentioned. Participants' thinking appears to have shifted discretely when quantum mechanics was explicitly mentioned in the questions about generic measurement uncertainty. That is, participants' ideas about uncertainty were more polarized when considering uncertainty generically than in response to hypothetical data sets. Future work should evaluate the role of framing and resources as students describe measurement uncertainty in different contexts. If student reasoning differs between contexts (classical vs. quantum and generically vs. in response to data), this would have implications for assessments of student thinking about measurement uncertainty. 

Our analysis also sheds light on distinctions between student thinking about accuracy and precision in the classical and quantum contexts. Students frequently drew on measurement limitations in discussing both accuracy and precision across both contexts. This supports existing work that students tend to conflate sources of uncertainty and errors or mistakes~\cite{HolmesBFY2015, Evangelinos2002}. Furthermore, when comparing two data sets, participants tended to give similar answers about both the ball-drop and single-slit experiments. This is in stark contrast to the differences between the contexts with regard to precision and generic uncertainty. When considering a comparison between a data set and a hypothetical theory calculation, participants similarly mentioned measurement limitations with the same frequency for both experiments. However, participants drew on physical limitations, usually to question the completeness of the theory, with similar frequency as measurement limitations in the ball-drop experiment, while for the single-slit experiment, few participants drew on physical limitations. These responses are consistent with previous work indicating that upper-level students see theory both as validating and being validated by experiment~\cite{Hu2018,Hu2017}, though our analysis notably suggests that students are less likely to see experiment as validating theory in a quantum context.

\section{Conclusion}

This study shed light on student thinking about measurement uncertainty in classical and quantum contexts, drawing on the Modeling Framework for Experimental Physics~\cite{zwickl2015model} to code student ideas about sources of measurement uncertainty. In addition to indicating that students think about measurement uncertainty quite differently across physics contexts and types of uncertainty, the results suggest that student thinking about uncertainty may be sensitive to cueing from questions. These results will inform the design of future assessments to probe ideas about sources of measurement uncertainty in a larger and broader sample of students.

\acknowledgments{This work has been supported in part by the NSF under Grants No.~DUE-1808945 and No.~DUE-1809183. The authors would like to acknowledge the contribution of Martin Stein, who developed the protocol for and conducted the interviews used in this study.}

\bibliography{refs} 

\end{document}